\documentclass{IEEEtran}
\usepackage{cite}
\usepackage{amsmath,amssymb,amsfonts}
\usepackage{algorithmic}
\usepackage{graphicx}
\usepackage{textcomp}
\usepackage{bm,flushend}
\usepackage{comment}
\usepackage{xcolor,algorithm,algorithmic}
\def\BibTeX{{\rm B\kern-.05em{\sc i\kern-.025em b}\kern-.08em
    T\kern-.1667em\lower.7ex\hbox{E}\kern-.125emX}}
\begin{document}
\title{Adaptive Polynomial Chaos Expansion for Uncertainty Quantification and Optimization of Horn Antennas at SubTHz Frequencies}
\author{Aristeides D. Papadopoulos, \IEEEmembership{Member, IEEE}, Yihan Ma, Qi Luo,\\ and George C. Alexandropoulos, \IEEEmembership{Senior Member, IEEE}
\thanks{Part of this paper will be presented in the $18$-th European Conference on Antennas and Propagation, Glasgow, Scotland, 17--22 March 2024~\cite{Our_Eucap}.}
\thanks{A. D. Papadopoulos and G. C. Alexandropoulos are with the Department of Informatics and Telecommunications, National Kapodistrian University of Athens, Panepistimioupolis Ilissia,  15784 Athens, Greece (e-mail: \{aristep, alexandg\}@di.uoa.gr). }
\thanks{Y. Ma and Q. Luo are with the  School of Physics, Engineering, and Computer Science, University of Hertfordshire, Hatfield, AL10 9AB, UK (e-mail: \{y.ma20, q.luo2\}@herts.ac.uk).}
}

\maketitle

\begin{abstract}
Sub-terahertz (subTHz) antennas will play an important role in the next generations of wireless communication systems.  However, when comes to the subTHz frequency spectrum, the antenna fabrication tolerance needs to be accurately considered during the design stage. The classic approach to studying the average performance of an antenna design considering fabrication tolerances is through the use of the Monte-Carlo (MC) method. In this paper, we propose an adaptive polynomial chaos expansion (PCE) method for the uncertainty quantification analysis of subTHz horn antennas with flat-top radiation patterns. The proposed method builds a surrogate model of the antenna's response to electromagnetic (EM) excitation and estimates its statistical moments with accuracy close to the reference MC method, but with a much smaller computational complexity of roughly two orders of magnitude. Moreover, the surrogate model based on PCE can substitute full-wave EM solvers in producing samples for electromagnetic quantities of interest, resulting in significant computational efficiency gains during optimization tasks. To this end, we successfully combined PCE with the particle swarm optimization method to design the free parameters of a horn antenna at $95$ GHz for a flat-top gain.  
\end{abstract}


\section{Introduction}
\label{sec:introduction}
\IEEEPARstart{T}{he} increasing demands on high data rate applications have been motivating wireless communications in high frequencies, such as the frequency range $2$ (FR2) in the fifth generation (5G) new radio (NR) and the envisioned sub-terahertz (subTHz) and THz bands~\cite{Guo2022} for the upcoming 5G-Advanced and the future sixth generation (6G) of wireless networks. For THz transceivers, microstrip antennas~\cite{Lu2022}, dielectric resonator antennas, waveguide antennas~\cite{Mon2019}, and lens antennas~\cite{Hao2017} are being developed, while there also lately exists significant interest for THz metasurfaces with tunable reflection properties~\cite{Eucap_Terrameta}. In general, an antenna's size is proportional to the wavelength of the frequency that the antenna operates. This implies that, when comes to subTHz and THz frequency spectra, the dimensions of the antennas become small and the fabrication tolerance must be considered during the design stage~\cite{Hao2018}. Although high-precision fabrication technologies~\cite{Diana2016,yang2022} are available, it would be more cost-effective to fabricate antennas and metasurfaces using conventional technologies, such as printed circuit board (PCB) and computer numerical control (CNC) milling. 

To study the effect of fabrication tolerance on the performance of THz devices, the classical approach is to vary the design parameters in electromagnetic (EM) simulators~\cite{xiang2018}. The Monte Carlo (MC) method is a widely known method for studying uncertainty. In particular, the EM simulator is executed a large number of times to derive samples of the structure's performance measure. Then, directly from these samples, the statistical moments of the performance measure are computed. If there is a large number of design variables, thousands of EM simulator runs are required for the MC method, which is time-consuming and requires lots of computation resources. On the contrary, the polynomial chaos expansion (PCE) method would require a few dozens of them~\cite{Xiu2002}. PCE builds a surrogate model of a system's performance measure. This model provides a series expansion of the quantity of interest (QoI), which is specified by deterministic coefficients and stochastic polynomial basis functions. The size of the series is reduced by considering only terms of low-order interactions. Then, the most significant series coefficient terms are estimated via compressed sensing (CS) tools, e.g., the Orthogonal-Matching-Pursuit (OMP) algorithm~\cite{tropp2007}. With OMP, a few deterministic model evaluations are required and an accurate PCE surrogate is built with its deterministic coefficients forming a sparse vector. In~\cite{papadopoulos2022}, an adaptive PCE method was proposed that considers multiple PCE models of increasing polynomial complexity and optimally chooses one of the lowest errors avoiding overfitting.  

The PCE method has been applied in different forms in the past for uncertainty quantification in antenna design. In~\cite{Boeykens2014}, the resonance frequency of a curved textile antenna was statistically assessed under random variations of the radius of curvature using the PCE model. This model focused on a single random design variable, namely the curvature, and the estimation of the PCE coefficients involved Galerkin projection for the PCE expansion and Gaussian quadrature for integral calculations. Although the presented method demonstrated increased efficiency and accuracy compared to MC, it was recognized that Galerkin projection encounters the curse of dimensionality as the number of random design variables increases. Furthermore, the method did not account for the sparsity in the PCE coefficients or the fact that the most significant interactions in the PCE typically have low polynomial orders. In~\cite{Wilke2017}'s study, the performance of a satellite antenna at Ka-band with $19$ uncertain design variables was analyzed using the PCE method. The sparsity in the PCE coefficients was addressed by employing least angle regression in conjunction with a  least-squares method. However, higher-order polynomial interactions were kept without an adaptive scheme, resulting in an algorithm with reduced robustness, as compared with the adaptive PCE method of~\cite{papadopoulos2022}. A quad-mode antenna with $4$ uncertain design variables was analyzed via PCE in~\cite{Klink2021}. Various solvers, including non-sparse ones like the least-squares method, were employed to determine the PCE expansion coefficients. However, again, no reduction of high-order polynomial terms was considered. Specifically, different sampling schemes were utilized, but some were inconsistent with the Wiener-Askey exponential converging PCE scheme. Ultimately, $35$ required full-wave EM samples were obtained. 

The available open technical literature also introduces various approaches based on surrogate modeling to expedite EM optimization. In~\cite{Zhang2023}, the authors employed machine learning coupled with data augmentation for antenna optimization, utilizing a random forest regressor as a surrogate model. The presented method relied on an adaptive data augmentation algorithm, invoking a full-wave EM solver for additional samples when necessary. The method was successfully applied to optimize a circularly polarized omnidirectional base station antenna with $4$ design variables, and for this goal, $81$ full-wave EM samples were utilized. In a different study in~\cite{Liu2022}, the differential evolution optimization method was combined with a surrogate model based on a Bayesian neural network. That network was trained through adaptive design variable selection using the lower confidence bound method. The presented technique demonstrated success in challenging antenna optimization cases, utilizing $924$ full-wave EM simulations for an antenna with $10$ geometric design variables.

In this paper, we deploy for the first time the adaptive PCE method of~\cite{papadopoulos2022} to perform uncertainty quantification analysis of horn antennas with flat-top radiation patterns, considering the fabrication tolerance of the conventional CNC milling. Through extensive performance evaluations, it is demonstrated that the proposed framework can effectively reduce the computational time of uncertainty quantification analysis, while exhibiting similar accuracy with conventional MC. In addition, we apply the presented adaptive PCE method as a generator of quantities of interest (QoI) samples in collaboration with the particle swarm optimization (PSO) algorithm~\cite{Kennedy1995} to facilitate highly efficient optimization of the gain for the flat-top horn antenna. Compared with previous works, our proposed method stands out as superior, providing adaptability and increased robustness compared to~\cite{Boeykens2014}~-~\cite{Klink2021}, keeping the most significant low-order polynomial terms in the PCE. Addressing a horn optimization problem with $9$ geometric design variables, we employed only $100$ full-wave EM simulations. Additionally, our method yielded improved uncertainty quantification analysis results compared to other state-of-the-art surrogate-assisted antenna optimization methods.

The remainder of this paper is organized as follows. Section~\ref{sec:pce} overviews the theory of the adaptive PCE method, while Section~\ref{sec:horn} presents the configuration of the horn antenna used as a case study. Section~\ref{sec:results} includes the uncertainty quantification  and optimization results, whereas Section~\ref{sec:conclusion} concludes the paper. 

\textit{Notations:} Lower and upper case boldface letters denote vectors and matrices, respectively. $\lVert\cdot\rVert_q$ and $\odot$ indicate the $q$-norm of a vector and the element-wise product, respectively. ${\rm Tr}\{\cdot\}$ returns the trace of a matrix, $\mathbb{N}_0$ is the set with the natural numbers including zero, and $\mathbb{R}$ is the real number set.

\section{Adaptive Polynomial Chaos Expansion}\label{sec:pce}
In this section, we commence with a description of the main features of the PCE methodology and then present an algorithmic approach for the computation of the adaptive PCE coefficients. We also discuss the benefits from using the PCE of a QoI to efficiently generate samples for it as well as to perform estimation of its statistical functions and moments.

\subsection{Preliminaries}
Consider a QoI $u$ that depends on a number $d$ of independent random variables included in the vector $\bm{\xi} \triangleq \left[\xi_{1}, \xi_{2}, \ldots, \xi_{d} \right]\in \mathbb{R}^{d}$. The PCE surrogate model of this QoI $u$ has the following mathematical form~\cite{Xiu2002}:
\begin{equation}\label{PCE}
u\left( \bm{\xi} \right) \cong \sum_{n}c_{\bm{\alpha}_n}\Psi_{\bm{\alpha}_n}\left( \bm{\xi} \right), 
\end{equation} 
where each $n$-th term ($n=1,2,\ldots$) of the series depends on a distinct feasible vector $\bm{\alpha}_n\triangleq\left[\alpha_{1n}, \alpha_{2n}, \ldots, \alpha_{dn} \right] \in \mathbb{N}_0^{d}$ and $c_{\bm{\alpha}_n}$ represents the PCE's deterministic coefficients of each $n$-th basis function $\Psi_{\bm{\alpha}_n}$, which is defined as follows: 
\begin{equation}
\Psi_{\bm{\alpha}_n}\left( \bm{\xi} \right)\triangleq\prod_{i=1}^{d}\psi_{\alpha_{in}} \left( \xi_{i} \right). \label{BasisFunctions}
\end{equation}
According to this expression, each $\alpha_{in}$ component ($i=1,2,\ldots,d$) of each $n$-th feasible $\bm{\alpha}_n$ indicates the degree of the univariate  function $\psi_{\alpha_{in}}\left(\cdot\right)$. In this paper, we assume that $\xi_{i}$'s are uniformly distributed and the functions $\psi_{\alpha_{in}} \left(\cdot\right)$'s are Legendre polynomials~\cite{Legendre}. This assumption has been proved to lead to an exponentially convergent PCE expansion~\cite{Xiu2002}; it is though noted that there exist  different choices of the latter parameters that can guarantee exponential convergence. In addition, the set containing the multiple indices $\bm{\alpha}_n$'s is chosen such that it forms the anisotropic-hyperbolic index set:
\begin{equation}
\mathbb{A}_{q,\bm{w}}^{d,p}\triangleq\left\{ \bm{\alpha}_n: \left[ \sum_{i=1}^{d}\left(w_{i}\alpha_{in}\right)^{q} \right] ^{1/q} \leq p    \right\}, \label{indices}
\end{equation}
where $\bm{w}\triangleq[w_1,w_2,\ldots,w_d]\in \mathbb{R}^{d+}$ is a weight vector that induces anisotropy (i.e., polynomials combined with larger weights are not favored)~\cite{blatman2009}, $q\in(0,1]$, and $p$ is the maximum of the polynomial orders present in the PCE series. It can be observed that, when $\bm{w}=\bm{1}$, the weighted-anisotropic index set becomes the more commonly used, but less robust, hyperbolic index set  $\mathbb{A}_{q}^{d,p}$ \cite{papadopoulos2022}.
The hyperbolic truncation scheme reduces the number of polynomials used in the PCE expansion since higher-order polynomials are penalized. This choice works well for PCE series and it is accurate in most cases. As it is known in the literature, the latter holds due to the fact that interactions of low-order polynomials dominate the description of many practical problems. 

The weights in the $\mathbb{A}_{q,\bm{w}}^{d,p}$ definition, which favor anisotropy in the selection of the polynomial order, provide an additional degree of freedom that makes the index truncation scheme more robust. This is achieved by relating these weights with the total Sobol indices $S_{i}^{T}$'s. These indices are defined as the fraction of the variance of each random variable $\xi_{i}$ over the total variance $\sigma^{2}\left(u\right)$ of the QoI $u$~\cite{Crestaux2009}. For the considered PCE case, each $S_{i}^{T}$ can be obtained as a function of the PCE deterministic coefficients, i.e.:
\begin{equation}\label{TS}
S_{i}^{T} = \sum_{n}c_{\bm{\alpha}_n}^{2}\sigma^{-2}\left(u\right)\int_{\mathbb{R}^{d}} |\Psi_{\bm{\alpha}_n}\left( \bm{\xi} \right)|^2{\rm w}\left( \bm{\xi} \right){\rm d}\bm{\xi},
\end{equation}
where ${\rm w}\left( \bm{\xi} \right)$ represents the probability distribution function (PDF) of $\bm{\xi}$ and each $n$-th series term in \eqref{TS} depends on $\bm{\alpha}_n$ with $\alpha_{in}>0$ $\forall$$i$. In addition, each $w_{i}$ component of the weight vector $\bm{w}$ can be chosen as a function of $S_{i}^{T}$'s according to~\cite{blatman2009}, yielding the following expression:

\begin{equation}\label{w}
w_{i}=1+ \left(-S_{i}^{T}+\max_{j \in \{ 1,2,\ldots, d \} } S_{j}^{T} \right)\left(\sum_{j=1}^{d} S_{j}^{T}\right)^{-1}. 
\end{equation} 
The logic behind this update is that, as $S_{i}^{\left( T\right)}$ increases, $w_{i}$ decreases, and most likely,
the polynomial indices corresponding to $\xi_{i}$ will not be eliminated after applying the anisotropic truncation scheme of \eqref{indices}.

\subsection{CS-Based Computation of the PCE Coefficients}
It is essential for the PCE method to obtain the  $c_{{\bm\alpha}_n}$'s coefficients. As it is well-known from~\cite{blatmanPhD}, the \textit{sparsity of effects principle} holds for this method, which states that most of these coefficients are zero or almost zero. This implies that, if $\mathbf c$ is the vector containing all $N$ PCE coefficients, i.e., 
$\mathbf{c}\triangleq\left[c_{{\bm\alpha}_1},c_{{\bm\alpha}_2},\ldots,c_{{\bm\alpha}_N}\right]\in \mathbb{R}^{N}$, it should be sparse, and hence, CS methods can be deployed to estimate it~\cite{tropp2007}. 

Let the real-valued $M\times N$ dictionary matrix $\mathbf{\Phi}$ with its $(m,n)$-th element ($m=1,2,\ldots,M$ and $n=1,2,\ldots,N$) defined as $[\mathbf{\Phi}]_{m,n} \triangleq \Psi_{{\bm \alpha}_n}(\bm \xi^{(m)})$, where $\bm\xi^{(1)},\bm\xi^{(2)},\ldots,\bm{\xi}^{(M)}$ represent $M<N$ realizations of the random vector $\bm{\xi}$. The following CS problem can be formulated:  
\begin{equation}\label{minimization_problem}
{\mathbf{c}}^\ast \triangleq \mathop {\arg \min }\limits_{\mathbf{c}} {\left\| {\mathbf{c}} \right\|_0}\quad \textrm{subject to} \quad {\mathbf{\Phi c}} = {\mathbf{y}},
\end{equation}  
where $\mathbf{y}\triangleq\left[y(\bm \xi^{(1)}),y(\bm \xi^{(2)}),\ldots, y(\bm \xi^{(M)})\right]$ (also called measurement vector) and $\left\| {\mathbf{c}} \right\|_0$ provides the number of non-zero elements in $\mathbf c$.
Clearly, the solution of \eqref{minimization_problem} is the sparsest vector ${\mathbf{c}}^\ast$. Finding it though, has been shown to be infeasible since a search through all possible support sets needs to be performed. In this paper, we adopt the OMP method to approximate \eqref{minimization_problem}'s solution~\cite{tropp2007}. OMP is an iterative greedy algorithm aiming to find the columns of $\mathbf{\Phi}$ participating the most in the construction of the measurements contained in $\mathbf{y}$.
In every OMP iteration, a column-index set $\mathcal A$ is augmented and a residual $\mathbf{r}$ vector is calculated. The algorithm is initialized with the residual $\mathbf r^{(0)}$ set equal to $\mathbf{y}$ and the empty set $\mathcal{A}^{(0)}$. During each $k$-th iteration, $[\bm\Phi]_{:,n}^{\mathrm T}\mathbf{r}^{(k)}/ \left\| [\bm\Phi]_{:,n} \right\|_{2} $ ($[\bm\Phi]_{:,n}$ is the $n$-th column of $\bm{\Phi}$) is calculated, indicating the correlation of each column of $\bm \Phi$ (excluding those already included in $\mathcal A^{(k-1)}$) with the residual, and the index corresponding to the maximum correlation is included to $\mathcal A^{(k)}$. Then, the new estimate of the sparse solution vector is obtained from the solution of the following least-squares problem:
\begin{equation}
{{\mathbf{c}}^{(k)}} = \mathop {\arg \min }\limits_{\mathbf{c}} {\left\| {{{\mathbf{\Phi }}_{{\mathcal A^{(k)}}}}{\mathbf{c}} - {\mathbf{y}}} \right\|_2},
\end{equation}
where ${{\mathbf{\Phi }}_{{\mathcal A^{(k)}}}}$ is the submatrix of $\mathbf{\Phi}$ that contains the columns with indices in $\mathcal A^{(k)}$. The last step of each $k$-th iteration is to update the residual as follows:
\begin{equation}
{{\mathbf{r}}^{(k)}} = {\mathbf{y}} - {{\mathbf{\Phi }}_{{\mathcal A^{(k)}}}}{{\mathbf{c}}^{(k)}}.
\end{equation}
The number $s$ of non-zero coefficients in $\mathbf{c}$ is pre-selected and the iteration index is set as $k = 1,2,\ldots,s$. Alternatively, OMP iterations may terminate when ${\left\| {\bm\Phi {\mathbf{c}}^{(k)} - \mathbf{y}} \right\|_2}$ becomes smaller than a given threshold $\epsilon$.

The estimation error of the latter OMP-based approximation, $\epsilon_{\mathbf{c}^{\ast}\left(p\right)}$, can be calculated via the leave-one-out cross validation method~\cite{tropp2007}. In particular, this error is given by:
\begin{equation}
\epsilon_{\mathbf{c}^{\ast}\left(p\right)}= CT\left(p,N_{p}\right) L\left(M_{X,p}\right),\label{eq:err}
\end{equation}
where $X$ denotes the set with the random sample points $x^{\left(i\right)}$ with $i=1,2,\ldots, N_{p}$ (i.e., the set used in the experimental study), where $N_{p}$ is the number of these points in the PCE of maximum polynomial order $p$, and $L\left(M_{X,p}\right)$ is the leave-one-out cross-validation error, which is calculated as follows~\cite{blatmanPhD}:
\begin{equation}
L\left(M_{X,p}\right)\triangleq\frac{1}{N_{p}-1}\sum_{i=1}^{N_{p}} \frac{QoI\left(x^{\left( i \right)}\right)-M_{X}\left(x^{\left( i \right)}\right)}{1-h_{i}}.\label{loo_err}
\end{equation}
In this definition, $QoI\left(x^{\left( i \right)}\right)$ is the QoI corresponding to the $x^{\left(i\right)}$ input, while $M_{X}\left(x^{\left( i \right)}\right)$ indicates the output of a PCE that is built from all the measurements $x^{\left( i \right)} \in X$. In addition, $h_{i}$ represents the $i$-th diagonal term of $\bm{\Phi}\left(\bm{\Phi}^\mathrm{T}\bm{\Phi}\right)^{-1}\bm{\Phi}^\mathrm{T}$. Finally, the correction term $CT\left(p,N_{p}\right)$ in~\eqref{eq:err} is designed to reduce the sensitivity of $L\left(M_{X,p}\right)$ in the overfitting regime (i.e., when $p$ increases), and is defined in~\cite{chapelle2002} as follows:
\begin{equation}
CT\left(p,N_{p}\right)\triangleq\left(1-\frac{p}{N_{p}} \right)^{-1}\left( 1+{\rm Tr}\left\{ \left( \bm{\Phi}^\mathrm{T}\bm{\Phi}  \right)^{-1}\right\}  \right).\label{CT}
\end{equation}	  

\begin{algorithm}[!t]
    \caption{Calculation of Adaptive PCE Coefficients}
    \label{alg:adaptive_PCE}
    \begin{algorithmic}[1]
        \renewcommand{\algorithmicrequire}{\textbf{Input:}}
        \renewcommand{\algorithmicensure}{\textbf{Output:}}
        \REQUIRE $\bm{w} = \bm{1}_{d}$ (hyperbolic truncation), $\epsilon$, $i=1,2,\ldots,d$, $p_{\min}$, $p_{\max}$, and $n=1,\ldots,N_{p}$.
        \ENSURE The PCE coefficients $\mathbf{c}^{\ast}_{f}$.
        \STATE Set $\epsilon_{\mathbf{c}^{\ast}\left(p_{\rm min}-1\right)}=\epsilon_{\mathbf{c}^{\ast}\left(p_{\rm min}-2\right)}=0$.
        \FOR{$p = p_{\rm min}, p_{{\rm min}+1},\ldots,p_{\rm max}$}
            \STATE Generate $\mathrm{A}_{p,\bm{w}}^{q,d}$ using \eqref{indices}.
            \STATE Calculate the OMP solution $\mathbf{c}^{\ast}\left(p\right)$ of \eqref{minimization_problem} or \eqref{Wminimization_problem} and compute the error $\epsilon_{\mathbf{c}^{\ast}\left(p\right)}$.
            \IF{$\epsilon_{\mathbf{c}^{\ast}\left(p\right)}<\epsilon$}
            \STATE Stop the algorithm and output $\mathbf{c}^{\ast}_{f} = \mathbf{c}^{\ast}\left(p\right)$.
            \ENDIF
            \IF{$\epsilon_{\mathbf{c}^{\ast}\left(p\right)}>\epsilon_{\mathbf{c}^{\ast}\left(p-1\right)}>\epsilon_{\mathbf{c}^{\ast}\left(p-2\right)}$ (overfitting)}
            \STATE Stop the algorithm and output $\mathbf{c}_{f}^{\ast}=\mathbf{c}^{\ast}\left(p-2\right)$.
            \ENDIF
            \STATE Compute the total Sobol indices using \eqref{TS}. 
            \STATE Update the weights $w_i$ $\forall$$i$ using \eqref{w}.
        \ENDFOR
        \STATE Set  $\mathbf{c}^{\ast}_{f}=\arg\min_{p}\{ \epsilon_{\mathbf{c}^{\ast}\left(p_{\min}\right)},\epsilon_{\mathbf{c}^{\ast}\left(p_{\min}+1\right)},\ldots,\epsilon_{\mathbf{c}^{\ast}\left(p_{\max}\right)} \}$.
    \end{algorithmic}
\end{algorithm}

The steps of the adaptive PCE method are summarized in Algorithm~\ref{alg:adaptive_PCE}.

To enhance the sparsity of the PCE coefficients and, thus, require less sampling points $G$ for a successful reconstruction, the PCE coefficients are found via a weighted $l_{0}$-minimization \cite{candes2008}. In particular, the optimization problem \eqref{minimization_problem} is rewritten in the following form:
\begin{equation}
{\mathbf{c}}^\ast \triangleq \mathop {\arg \min }\limits_{\mathbf{c}} {\left\| {\mathbf{Wc}} \right\|_0}\quad \textrm{subject to} \quad {\mathbf{\Phi c}} = {\mathbf{y}},\label{Wminimization_problem}
\end{equation}
where the matrix $\mathbf{W}$ is diagonal having the elements $w_{i}$ $\forall$$i=1,2,\ldots,N$ on its main diagonal. These weights are considered as free parameters which are chosen to improve the PCE coefficients reconstruction. In particular, if the solution of \eqref{minimization_problem} is solved by a particular algorithm, i.e. the OMP, then its corresponding weighted counterpart \eqref{Wminimization_problem} can be solved using $l_{max}$ successive applications of OMP, where at the $j$-th iteration the weights can be chosen as follows: 
\begin{equation}
w_{i}^{\left(j\right)}=\left(|c_{i}^{\ast\left(j\right)}|+\epsilon_{w}  \right)^{-1},\label{w_update}
\end{equation}
where $j=1,2,\ldots,l_{max}$,  $i=1,2,\ldots,N$, and $\epsilon_{w}$ is a small number used to avoid division by zero.
This expression eventually forces to zero the small-valued coefficients in the recovered $\mathbf{c}^{\ast}$ vector \cite{candes2008}, thus, enhancing sparsity in the solution. This fact leads effectively to a smaller number of required measurements $M$ for accurate reconstruction. The efficiency of this approach, which we henceforth call as weighted OMP (WOMP), will be validated through the uncertainty quantification analysis of a horn antenna in Section~\ref{sec:results}. 

\subsection{QoI Samples Generation and Statistical Estimation}
By using the PCE coefficients $\mathbf{c}_{f}^{\ast}$ computed via Algorithm \ref{alg:adaptive_PCE}, the PCE surrogate model for the QoI $u$ can be built according to \eqref{PCE}, as follows:
\begin{equation}
u\left( \bm{\xi} \right) \cong \sum_{n}c_{f,\bm{\alpha}_n}^{\ast}\Psi_{\bm{\alpha}_n}\left( \bm{\xi} \right).\label{PCEgen}
\end{equation}
It can be easily observed that the right hand side (RHS) of this expression is actually a sample generator of $u$, which, according to the PCE theory, can be exponentially convergent to the true $u$ values as more terms are considered in the PCE. Interestingly, the generation of $u$-samples via \eqref{PCEgen} is highly computationally efficient, since these samples are obtained through an analytical formula. Hence, \eqref{PCEgen} can be used to trivially generate large amounts of QoI samples, which can be further used to approximate the PDF of $u$, and consequently, any order of its statistical moments. Additionally, this expression can feed with large numbers of QoI samples any algorithm dealing with the QoI optimization over the design variables of a system, e.g., an antenna. This constitutes a great deal of computational saving, which would otherwise need, in the case of an antenna design, a full wave solver for the QoI generation, making the optimization most likely infeasible. 

\section{Horn Antennas with Flat-Top Radiation}\label{sec:horn}
\begin{figure} [!t]
\centering
\includegraphics[width=0.7\columnwidth]{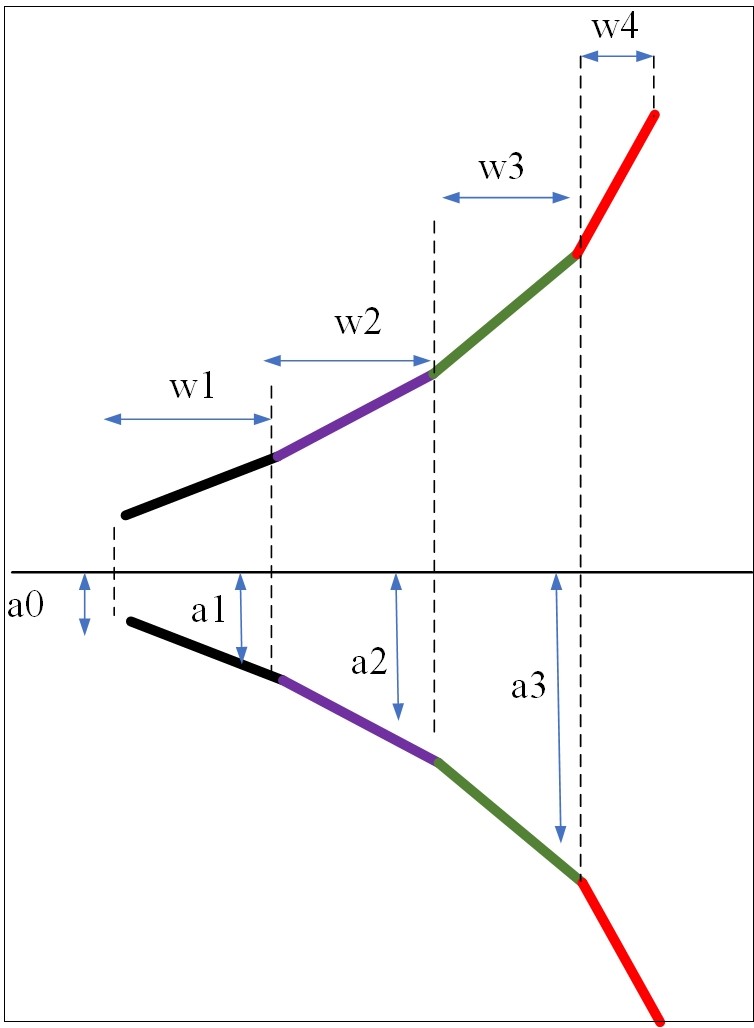}
\caption{Geometry and design parameters of the horn antenna in \cite{wang2019} operating at $95$ GHz, which is realized as a cascade of multiple linear circular waveguide sections.}
\label{HA}
\end{figure}

Horn antennas have been widely used as the feed for reflectors, reflectarrays, and transmitarrays~\cite{Imbriale2012}. Such a horn-made feed with a flat-top radiation pattern is highly desirable for reflectors since it can reduce the spillover and improve the illumination efficiency~\cite{mohamed2016}. One approach to realize a horn with a flat-top radiation pattern is to cascade multiple linear circular waveguide sections~\cite{wang2019}. Such a horn antenna configuration is illustrated in Fig.~\ref{HA}.

\begin{figure} [!t]
\centering
\includegraphics[width=0.9\columnwidth]{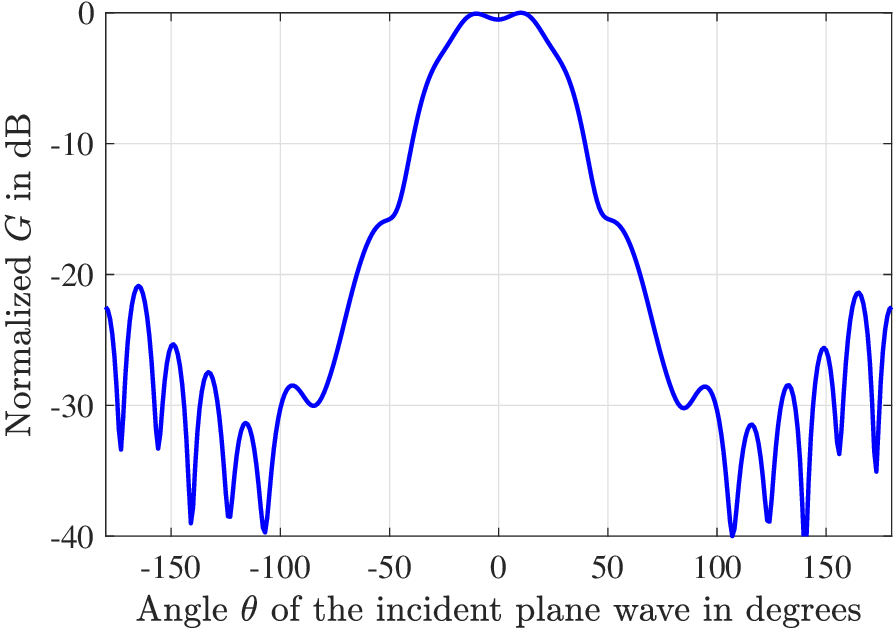}
\caption{The EM-simulated flat-top radiation pattern of the subTHz horn antenna shown in Fig.~\ref{HA} with the parameters $a_{0}=1.2$, $a_{1}=1.5$, $w_{1}=1.5$, $a_{2}=2.3$, $w_{2}=1.8$, $a_{3}=4.55$, $w_{3}=1.6$, $w_{4}=1.35$, and $a_{4}=6.85$ (all in mm) operating at $95$ GHz.}
\label{RP_horn}
\end{figure}

Figure~\ref{RP_horn} demonstrates the Ansys HFSS simulation results for the flat-top radiation pattern of a horn antenna with the parameters (all in mm) $a_{0}=1.2$, $a_{1}=1.5$, $w_{1}=1.5$, $a_{2}=2.3$, $w_{2}=1.8$, $a_{3}=4.55$, $w_{3}=1.6$, $w_{4}=1.35$, and $a_{4}=6.85$ operating at $95$ GHz. In the following section, we will deploy the adaptive PCE Algorithm~\ref{alg:adaptive_PCE}, in conjunction with EM simulations, to perform uncertainty quantification analysis considering the fabrication tolerance of CNC machining. To the best of our knowledge, for a medium tolerance class with a linear dimension from $6$ mm to $10$ mm, the tolerance is $\pm0.2$ mm, while with a linear dimension smaller than $6$ mm, the tolerance becomes $\pm0.1$ mm. Our goal with the proposed PCE-based analysis is to investigate how the fabrication tolerance affects the radiation pattern of a horn-made feed.

\section{Simulation Results and Discussion}\label{sec:results}
In this section, we present numerical results for the uncertainty quantification analysis of a subTHz horn antenna using the presented adaptive PCE method in Section~\ref{sec:pce}, which are also compared with equivalent ones obtained from MC simulations. We have focused on the gain $G$ of the subTHz horn antenna, whose geometry is illustrated in Fig.~\ref{HA}, as the QoI metric, and considered uncertainty in the antenna's nine optimized design parameters $[a_{0},a_{1},w_{1},a_{2},w_{2},a_{3},w_{3},w_{4},a_{4}]$ with respective values $[1.2,1.5,1.5,2.3,1.8,4.55,1.6,1.35,6.85]$, all in mm, for the maximum gain. It was assumed that the antenna is fabricated via the CNC method, and considering the limitation of this technology, the parameter $a_{4}$ varies by $\pm 0.2$~mm, while the other parameters can vary by $\pm 0.1$~mm. 

\subsection{First- and Second-Order Moments}
The proposed adaptive PCE method has been applied to compute the PCE coefficients via~\eqref{PCEgen}, which were first used to estimate the mean value and standard deviation (std) of the QoI $G$, respectively, as follows: 
\begin{align}
\mu_{G} &= c_{\mathbf{0}_d}\label{mu},\\
\sigma_{G}^{2} & = \sum_{n=1}^{\mathcal{N}} c_{\bm{\alpha_{n}}}^{2} \int_{\mathbb{R}^{d}} \left|\Psi_{\bm{\alpha}_n}\left( \bm{\xi} \right)\right|^2{\rm w}\left( \bm{\xi} \right){\rm d}\bm{\xi}\label{sigma}
\end{align}
where $\mathbf{0}_d$ is a vector with $d$ zeros and $\mathcal{N}$ denotes the cardinality of the set $ \mathbb{A}_{q,\bm{w}}^{d,p} - \{\mathbf{0}_d\}$ where all $\bm{\alpha_{n}}$'s belong to.
\begin{figure}[!t]
\centering
\includegraphics[width=0.9\columnwidth]{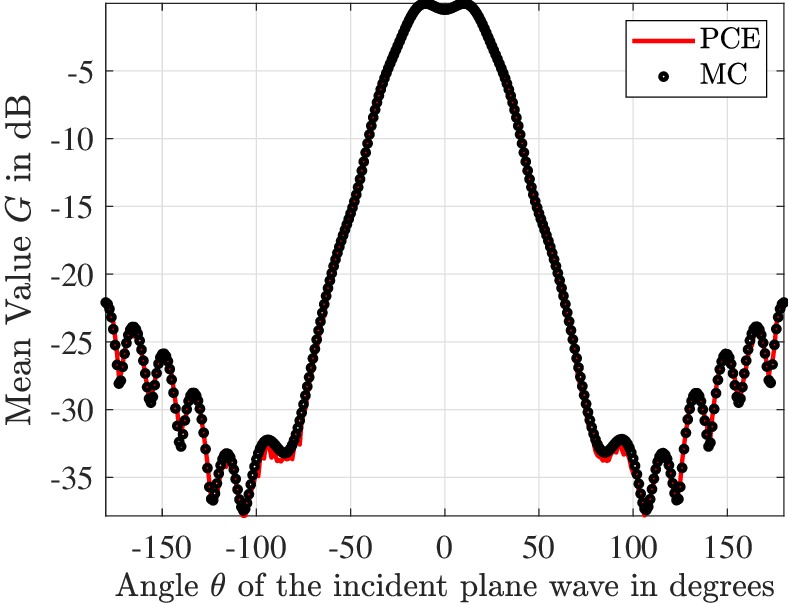}
\caption{The parameter $\mu_G$ versus the angle of radiation for the PCE method with $p_{\rm max}=4$ and $M=25$. Respective results via the MC method are also included.}
\label{mean}
\end{figure}

\begin{figure}[!t]
\centering
\includegraphics[width=0.9\columnwidth]{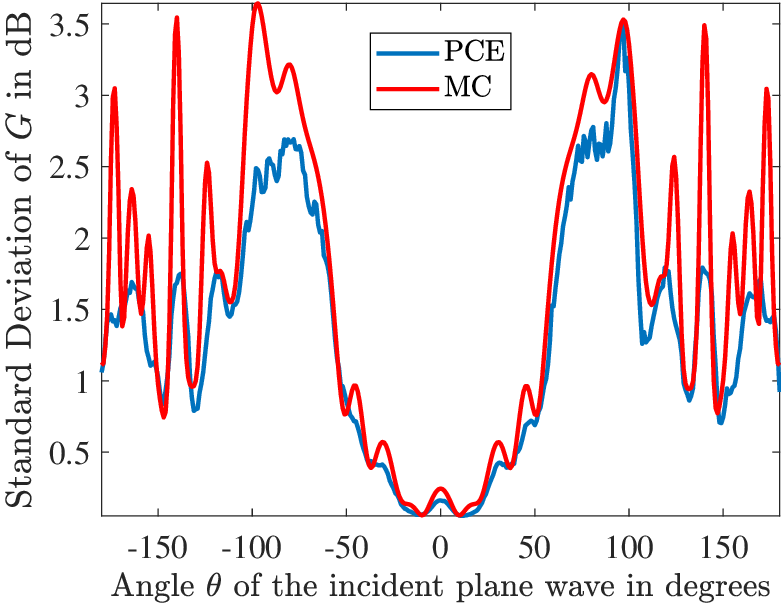}
\caption{The parameter $\sigma_G$ versus the angle of radiation for the PCE method with $p_{\rm max}=4$ and $M=25$. Respective results via the MC method are also included.}
\label{stda}
\end{figure}
In Fig.~\ref{mean}, the parameter $\mu_{G}$ is illustrated as a function of the angle of the incident plane wave at the frequency $95$ GHz, which was fixed for all numerical investigations in this paper. The expression \eqref{mu} was evaluated for the PCE method using $M=25$ independent $G$-samples and the maximum polynomial order $p_{\rm max}=4$, while, for the MC method, this metric was calculated directly via the mean value definition using $1000$ independent samples (almost two orders of magnitude more samples than PCE). As observed in the figure, there is excellent agreement between the two methods. It is noted that, to obtain every sample of $G$, a run of a full-wave EM finite element solver is required, which is significantly time-consuming, and thus, constitutes a major computational bottleneck. This figure actually showcases the power of the PCE method, which can serve as an accurate uncertainty quantification method that is much faster than the reference MC. As stated previously, for the optimal PCE convergence rate, the PCE random variable design samples need to follow the uniform distribution, while the MC samples are generated via the lattice hypercube method, where computation errors for $M$ samples reduce at a rate $1/M$ (which is better than the standard $1/\sqrt{M}$ rate of the MC method when samples are generated randomly). 
\begin{figure}[!t]
\centering
\includegraphics[width=0.9\columnwidth]{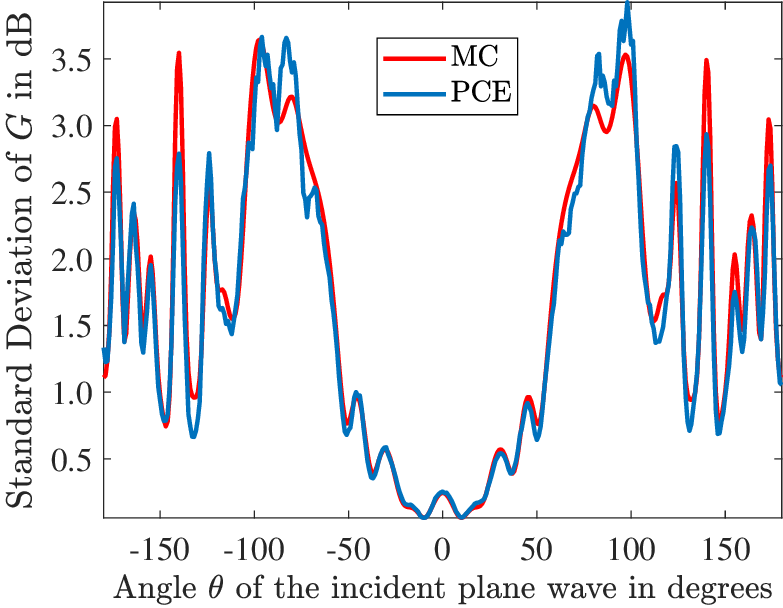}
\caption{The parameter $\sigma_G$ versus the angle of radiation for the PCE method with $p_{\rm max}=5$ and $M=50$. Respective results via the MC method are also included.}
\label{stdb}
\end{figure}

Parameter $\sigma_{G}$ is plotted versus the angle of the incident plane wave in Figs.~\ref{stda} and~\ref{stdb} for the two investigated methods. In Fig.~\ref{stda}, the same PCE parameter setting with Fig.~\ref{mean} has been used, while Fig.~\ref{stdb} includes PCE performance curves for $p_{\rm max}=5$ and $M=50$ independent antenna gain samples. It can be seen that there is a very good agreement between the two methods, with this agreement being improved in Fig.~\ref{stdb}, where a slightly higher maximum polynomial order and slightly more $G$-samples have been used. 

\begin{figure} [!t]
\centering
\includegraphics[width=0.9\columnwidth]{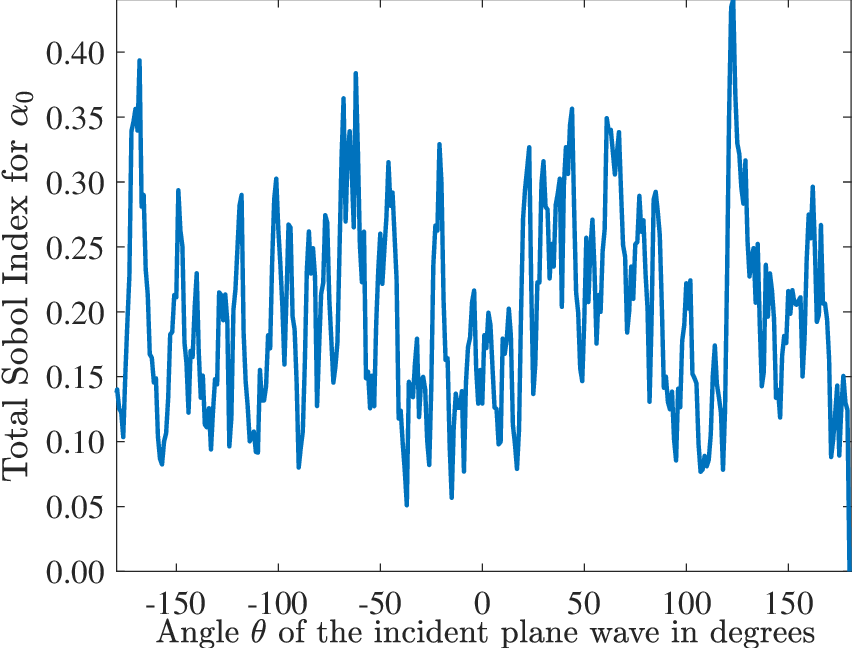}
\caption{The total Sobol index for the variable $a_{0}$ as a function of the angle of radiation.}
\label{sobol}
\end{figure}

According to the presented uncertainty quantification method in Section~\ref{sec:pce}, once the PCE coefficients are obtained, the Sobol indices can be calculated via expression $\eqref{TS}$ for every uncertain design variable. In Fig.~\ref{sobol}, the total Sobol index of the most uncertain variable $a_{0}$ (in the sense it varies the most with respect to its nominal value) is depicted as a function of the angle of the incident plane wave. This index quantifies, for every angle of incidence, the percentage of the $a_{0}$ variable in the total variance of the antenna gain $G$. Similarly, the total Sobol indices for all other design variables can be calculated, but are omitted here due to space limitations. 

\begin{figure} [!t]
\centering
\includegraphics[width=0.9\columnwidth]{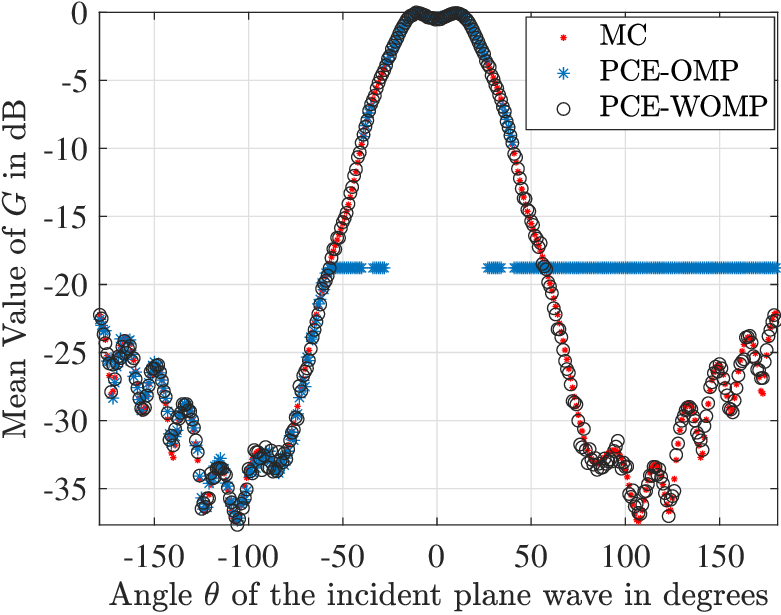}
\caption{The calculation of parameter $\mu_G$ with $M=20$ samples and for maximum polynomial order $p_{max}=5$. Successful and unsuccessful reconstruction when using the OMP and WOMP methods is also illustrated.}
\label{fig7}
\end{figure}

To improve the performance of the proposed PCE method in Algorithm~\ref{alg:adaptive_PCE}, we have replaced the OMP application in the corresponding algorithmic step (Step $4$) with the WOMP approach presented in~\eqref{Wminimization_problem} and~\eqref{w_update}. The corresponding performance evaluation is included in Fig.~\ref{fig7} for $M=20$ independent $G$-samples and the same maximum polynomial order $p_{max}=5$, where it can be observed that the WOMP solver reconstructs successively the PCE coefficients, generating average $G$-values that are in very good agreement with corresponding ones obtained via MC simulations. On the contrary, it is depicted that, for the same value of $M$ and maximum polynomial order, OMP fails. However, as it can be seen from Fig.~\ref{stdb}, when the number of independent $G$-samples increases to $M=50$, OMP yields successful reconstruction. This performance gap between OMP and WOMP can be explained from the fact that the weighted $l_{0}$-minimization algorithm in the latter is capable to increase the sparsity level of the reconstructed solution and, thus, WOMP requires smaller number of samples to achieve the same task.

\subsection{Statistical Functions}
A considerable number $N_{K}$ of samples for the horn antenna gain variable $G$ can be efficiently generated through the proposed adaptive PCE method, via~\eqref{PCEgen}, at a negligible computational expense. Subsequently, these samples can be directly used to construct a histogram of the $G$-samples, coupled with the application of a smoothing kernel, yielding an approximation for the PDF of variable $G$. In particular, the PDF of $G$, $f_{G}(\cdot)$, can be approximated as follows:
\begin{equation}
\hat{f}_{G}\left(y\right)=\frac{1}{N_{K}h_{K}}\sum_{i=1}^{N_{K}}K\left( \frac{y-y^{\left( i \right)}}{h_{K}}\right), \label{kernel_pdf}
\end{equation}   
where $K\left(\cdot\right)$ represents a positive-definite function known as the kernel, while $h_{K}$ serves as the bandwidth parameter. A Gaussian kernel, which is characterized by the standard normal PDF, can be chosen, and for this kernel, the optimal bandwidth is given by the formula:
\begin{equation}
h_{K}^{\ast}=\left(\frac{2}{3N_{K}}\right)^{1/5} \mbox{min}\left(\hat{\sigma}_{R}, \widehat{iqr}_{R}  \right),\label{opt_h}
\end{equation}   
where $\hat{\sigma}_{R}$ and $\widehat{iqr}_{R}$ denote the empirical standard deviation and interquartile range, respectively. As highlighted in~\cite{blatmanPhD}, the optimality of $\hat{\sigma}_{R}$ lies in its ability to minimize the $\ell_{2}$ approximation error $\|f_{G}(y)-\hat{f}_{G}(y)\|_{2}$.
\begin{figure} [!t]
\centering
\includegraphics[width=0.9\columnwidth]{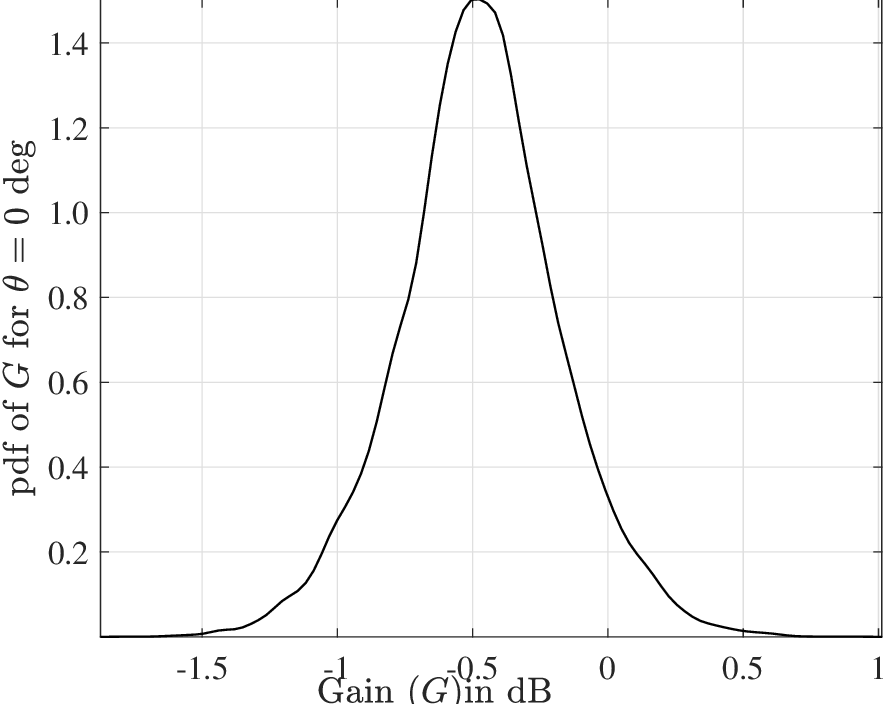}
\caption{The PDF of the antenna gain $G$ at the angle $\theta=0^{\circ}$ via the proposed adaptive PCE method in~\eqref{PCEgen}.}
\label{fig8}
\end{figure}

In Fig.~\ref{fig8}, the PDF calculated from PCE-based $G$-samples is plotted for the angle $\theta=0^{\circ}$ of the incident plane wave. It is noted that the PDF for any other angle value $\theta$ can be obtained in a similar way. In fact, in this way, any order probabilistic moment of $G$ can be calculated. This happens because expression~\eqref{PCEgen} is actually a $G$-sample generator which can be also used for optimization tasks in conjunction with an appropriately designed optimization algorithm.  

\begin{figure} [!t]
\centering
\includegraphics[width=0.9\columnwidth]{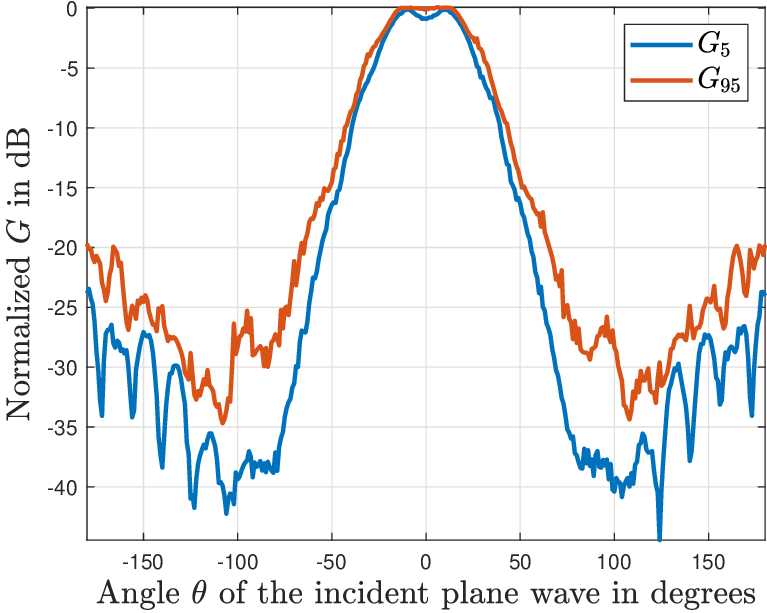}
\caption{The $5$-th ($G_5$) and $95$-th percentiles ($G_{95}$) of the antenna gain $G$ as calculated via the proposed adaptive PCE method in~\eqref{PCEgen}.}
\label{fig9}
\end{figure}

Another pertinent statistical metric is the $P$-th percentile of reflection with $0 \leq P \leq 100$, which can be efficiently computed via the proposed PCE method. By denoting each $P$-th percentile of $G$ as $G_{P}$, this metric signifies that $P\%$ of the observations of $G$ are smaller than $G_{P}$. After determining the PCE coefficients $\mathbf{c}_{f}^{\ast}$ in~\eqref{PCEgen}, a sufficiently large number $N$ of $G$-samples can be generated analytically with negligible computational cost. Subsequently, the $P$-th percentile $G_{P}$ of $G$ can be computed using the nearest-rank method. According to this method, the gain samples $G_{i}$, with $i=1,2,\ldots,N$, are first arranged in ascending order to form the set of sorted reflection observations $G_{s,i}$ with $i=1,2,\ldots,N$. Consequently, $G_{P}$ is determined as $G_{s,n}$ where $n$ is the smallest integer greater than or equal to $PN/100$. The $5$-th and $95$-th percentiles of $G$ as calculated via the proposed adaptive PCE method are depicted in Fig.~\ref{fig9}.

\subsection{Optimization of the Horn Antenna Gain}
The proposed adaptive PCE method via expression~\eqref{PCEgen} has been used as a generator of the QoI $G$-samples with the goal to maximize the gain of the horn antenna in Fig.~\ref{HA}. For this optimization, we have considered the PSO method~\cite{Kennedy1995} with the following cost function for a particle's solution $\mathbf{x}$:  
\begin{equation}
L\left( \mathbf{x}\right)=\sum_{m=1}^{5}\sum_{\theta \in R_{m}}\left|G_{\text{I}}\left(\mathbf{x},\theta\right)-G_{\text{PCE}}\left(\mathbf{x},\theta \right)\right|^4,\label{cost}
\end{equation}
where $G_{\text{I}}$ denotes the ideal antenna gain and the angle regimes $R_{m}$ with $m=1,2,3,$ and $4$ defined as $R_{1}: -180^{\circ}\leq \theta <-100^{\circ}$, $R_{2}: -100^{\circ}\leq \theta <-20^{\circ}$, $R_{3}: -20^{\circ}\leq \theta <20^{\circ}$, $R_{4}: 20^{\circ}\leq \theta <100^{\circ}$, $R_{5}: 100^{\circ}\leq \theta \leq 180^{\circ}$, which was combined with PCE, as described in the Appendix. This optimization approach was then deployed for the antenna gain maximization and the obtained optimum design variables are included in Table~\ref{table:1}. This table also lists the optimum design variables of the scheme presented in~\cite{wang2019}. As observed, the values for all antenna parameters with the proposed PSO-PCE design are very close to the ones via the benchmark scheme. 

\begin{table}[t!]
\centering
\caption{Optimized Horn Antenna Parameters via the Proposed PSO-PCE Approach and the Approach in \cite{wang2019}.}
\begin{tabular}{ |c|c| c|} 
\hline
Parameter (mm) & Proposed PSO-PCE & \cite{wang2019}  \\ \hline\hline
$a_{0}$ &1.227 & 1.2  \\ 
 $a_{1}$&1.482 & 1.5  \\ 
 $w_{1}$ &1.646 & 1.5  \\ 
 $a_{2}$&2.161 & 2.3  \\ 
 $w_{2}$&1.818 & 1.8  \\ 
 $a_{3}$& 4.846& 4.55  \\ 
 $w_{3}$&1.459 & 1.6  \\ 
 $w_{4}$&1.276 & 1.35  \\ 
 $a_{4}$&6.165 & 6.85   \\
 \hline
\end{tabular}
\label{table:1}
\end{table}
\begin{figure} [!t]
\centering
\includegraphics[width=0.9\columnwidth]{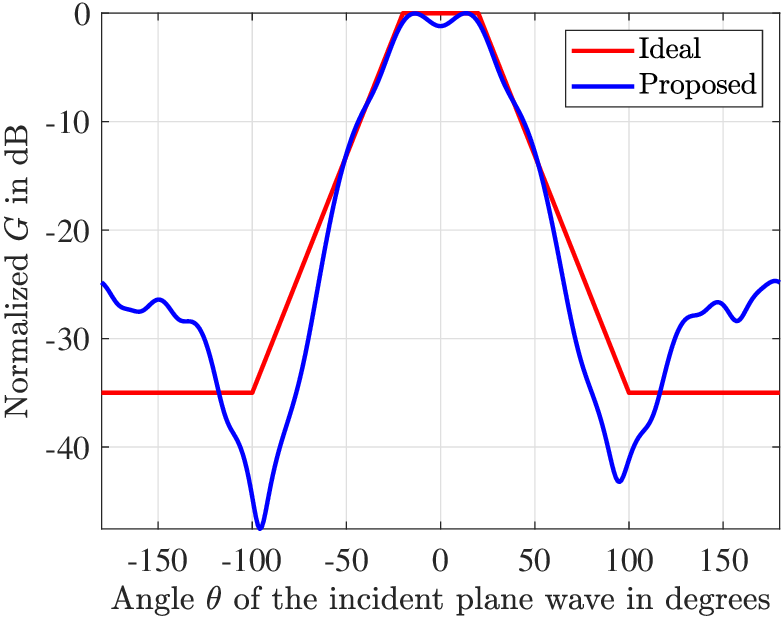}
\caption{The ideal and optimized, via the proposed PSO-PCE in the Appendix that deploys Algorithm~\ref{alg:adaptive_PCE}, horn antenna gain $G$.}
\label{fig10}
\end{figure}
Figure~\ref{fig10} illustrates the normalized horn antenna gain $G_{\text{PCE}}$ for the optimal design variables in Table~\ref{table:1} with the proposed PSO-PCE approach in comparison with the $G_{\text{I}}$ values. It can be observed that the flat top of the horn antenna with the proposed PSO-PCE approach within the range of angles $-50^{\circ}\leq \theta \leq 50^{\circ}$ closely aligns with the ideal case $G_{\text{I}}$. More importantly, this alignment is nearly identical to the gain depicted in Fig.~\ref{RP_horn}. In addition, it needs to be noted that the proposed PCE-based optimization exhibits significantly higher efficiency, given that the gain $G$-samples (a few thousands in number) provided to the optimizer were analytically computed using~\eqref{PCEgen}, as opposed to relying on a full-wave EM solver (e.g., HFSS). The latter takes approximately $5$ minutes on an average personal computer of average computational capabilities. In the proposed PSO-PCE antenna gain optimization approach, the PCE coefficients were determined using the adaptive PCE Algorithm~\ref{alg:adaptive_PCE}, utilizing 100 $G$-samples which were obtained via HFSS for uniformly distributed design variables with a 10$\%$ variation around the parameters specified in~\cite{wang2019}. 

\begin{figure} [!t]
\centering
\includegraphics[width=0.9\columnwidth]{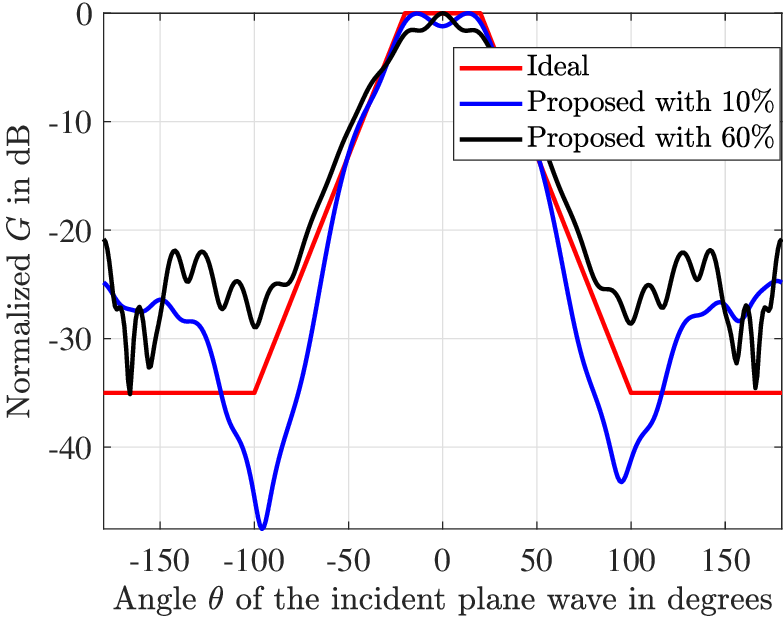}
\caption{The ideal and optimized, via the proposed PSO-PCE in the Appendix that deploys Algorithm~\ref{alg:adaptive_PCE}, horn antenna gain $G$ for the cases where the $G$-samples were provided to the PSO algorithm with $10\%$ (as in Fig.~\ref{fig10}) and $60\%$ variation around the parameters specified in \cite{wang2019}.}
\label{fig11}
\end{figure}

\begin{figure} [!t]
\centering
\includegraphics[width=0.9\columnwidth]{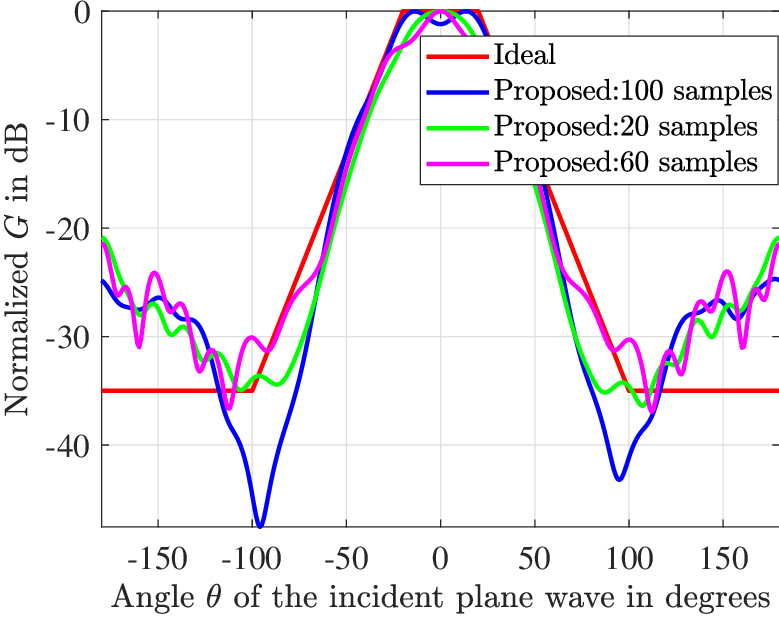}
\caption{The ideal and optimized, via the proposed PSO-PCE in the Appendix that deploys Algorithm~\ref{alg:adaptive_PCE}, horn antenna gain $G$ for the cases where the PCE model was constructed with $M=20$, $60$, and $100$ (as in Fig.~\ref{fig10}) number of $G$-samples.}
\label{fig12}
\end{figure}

In Fig.~\ref{fig11}, the performance of the proposed PSO-PCE antenna gain optimization approach is investigated for the case where the $G$-samples are generated solely via the proposed PCE model. In particular, the samples encompassed variations of $10\%$ or $60\%$ around the parameters specified in~\cite{wang2019}. As depicted, for the case of $60\%$ variation, the gain obtained from the proposed optimization closely aligns with the ideal one. This underscores the robust generalization properties of the proposed PCE model as a generator of $G$-samples, given its construction based on samples with only a $10\%$ variation around the reference parameters from~\cite{wang2019}. Finally, in Fig.~\ref{fig12}, the same results are demonstrated for the case where the $G$-samples are generated via the proposed PCE model in~\eqref{PCEgen}, and in particular, for $M=20$, $60$, and $100$ $G$-samples. It can be observed that the most accurate (closer to the ideal case) model in the region of interest $[-50^{\circ}, 50^{\circ}]$ (flat-gain) is the model created by $100$ $G$-samples. Notably, even with as few as $M=20$ samples, the resulting PCE model, coupled with the PSO algorithm, yields an optimized gain that is remarkably close to the ideal one. Overall this behavior underscores the impressive computational efficiency of the proposed PCE method in optimization endeavors.

\section{Conclusion}\label{sec:conclusion}
In this paper, an adaptive PCE approach for the efficient stochastic analysis of subTHz horn antennas was presented. The proposed method was successfully applied for the uncertainty quantification of such an antenna operating at $95$ GHz and realizing a flat-top radiation pattern with random geometric design variables. The presented numerical results showcased that the developed method is very accurate in the calculation of the QoI's mean and standard deviation values for the flat-top radiation pattern of the antenna (angle-band of interest), and sufficiently accurate outside of it. Indicatively, only $50$ samples in the random variable design space were used with the presented PCE method for uncertainty quantification, as compared to $1000$ samples used by MC. Moreover, by modifying the OMP algorithm used in the presented adaptive PCE method to a weighted CS version, the required PCE samples were reduced from $50$ to $20$. Recall that, for every QoI sample, a full-wave EM finite element solver needs to be executed. This signifies the immense saving in computation time of the PCE versus the standard MC method. In addition, the calculated PCE coefficients provide directly, for every frequency, the total Sobol indices for every uncertain design variable, which is valuable information for antenna design applications quantifying QoI's variance. 

After having estimated the PCE coefficients, large numbers of QoI samples can be generated analytically with insignificant computational cost. We used the presented PCE to directly calculate the PDF of the horn antenna (with the potential to calculate its statistical moments of any order) and its $5$-th and $95$-th percentiles. In addition, the PCE-based generator was deployed to obtain samples of the antenna gain which were used to feed the PSO algorithm with the goal to optimize the antenna design parameters. The extensive performance evaluation of the proposed PSO-PCE approach showed that the optimized horn design parameters are approximately the same with the optimum ones obtained via the state-of-the-art. This proves the potential of PCE, not just for uncertainty quantification analysis, but also for computationally efficient and accurate design optimization. In the future, we plan to apply the proposed PCE-based framework to other subTHz antenna designs and multi-functional metasurfaces~\cite{RISoverview2023}.

\section*{Acknowledgements}
This work has been supported by the UK Royal Society with grant number ${\rm IES}\! \ \!{\rm R}2\! \ \!212064$ and the Smart Networks and Services Joint Undertaking (SNS JU) project TERRAMETA under the European Union's Horizon Europe research and innovation programme under Grant Agreement No $101097101$, including top-up funding by UK Research and Innovation (UKRI) under the UK government's Horizon Europe funding guarantee.

\appendix
PSO constitutes a computationally efficient optimization technique drawing inspiration from the mathematical representation of a swarm of birds, which is utilized to determine the global solution of optimization problems. In the considered Horn antenna design problem, the candidate solutions reside within a space of $d=9$ dimensions. Within this $d$-dimensional space, each possible solution represents a point or particle. The PSO algorithm~\cite{Kennedy1995} commences by initializing with $\tilde{N}$ uniformly distributed random particles which traverse the $d$-dimensional solution space. These particles possess positions $\mathbf{x}_{i}^{\left( k \right)}$ and velocities $\mathbf{v}_{i}^{\left( k \right)}$ (with $i=1,2,\ldots,N$) at each $k$-th iteration of the algorithm. The positions and velocities of each particle undergo updates, taking into account their behavior akin to a swarm of birds. Specifically, for each $i$-th particle, the following iteration rules hold:
\begin{eqnarray}
\mathbf{v}_{i}^{\left( k+1 \right)}&=&\tilde{w}_{i}\mathbf{v}_{i}^{\left( k \right)}+c_{i}r_{1}^{\left( k \right)}\left(p_{i}^{\left( k \right)} \mathbf{1}_d-\mathbf{x}_{i}^{\left( k \right)}   \right) \nonumber \\  & & +s_{i}r_{2}^{\left( k \right)}\left(p_{g}^{\left( k \right)} \mathbf{1}_d-\mathbf{x}_{i}^{\left( k \right)}   \right),  \label{v-update}\\
\mathbf{x}_{i}^{\left( k+1 \right)}&=&\mathbf{x}_{i}^{\left( k \right)} + \chi_{i}\mathbf{v}_{i}^{\left( k+1 \right)},\label{x-update}
\end{eqnarray}
where $\mathbf{1}_d$ is a $d$-dimensional vector with all ones and 0 $\leq$ $\chi_{i}$ $\leq$ 1 $\forall$$i$ represents the learning rate that consequently determines the extent to which the velocity influences the position of the $i$-th particle. The inertia constant $\tilde{w}_{i}$ governs how tightly each $i$-th particle turns, with smaller values leading to faster convergence and a more focused search in a limited solution space. In addition, the random numbers $r_{1}$ and $r_{2}$ in the latter expressions are uniformly distributed in $[0,1]$ and play a role in the update process. The cognitive constant $c_{i}$ signifies the intelligence of each $i$-th particle in pursuing its own local best solution $p_{i}$, where an increase in $c_{i}$ prompts the particle to move towards the best solution found. Moreover, the social constant $s_{i}$ influences the inclination of each $i$-th particle to pursue the global best solution $p_{g}$, which is the minimum $p_{i}$ across all particles. 

The success of the PSO method strongly depends on the loss or merit function $L\left(\cdot\right)$ that is set to assess the quality of each solution $\mathbf{x}_{i}$ represented by the $i$-th particle in the given optimization problem. In this paper, we use \eqref{cost} as the loss function for the horn antenna gain maximization objective. It is noted that typically the antenna gain samples, i.e., the $G$-samples, are determined through a full-wave solver for each $\mathbf{x}_{i}^{\left( k+1 \right)}$ via \eqref{x-update} at each $k$-th algorithmic iteration. Given that the computation of each $G$-sample takes approximately $5$ minutes, running the PSO becomes practically infeasible. To this end, it is crucial to recognize that the iterative solution points $\mathbf{x}_{i}^{\left( k+1 \right)}$ from \eqref{x-update} are essentially random vectors since they are initialized with uniform randomness. Moreover, the coefficients $r_{1}$ and $r_{2}$ in the PSO updates are also random. As a result, the gain samples $G\left( \mathbf{x}_{i}^{\left( k+1 \right)} \right)$ are stochastically generated, and their PDF can be accurately approximated by the convergent PCE method~\cite{Xiu2002}. \textit{This fact provides the rationale for replacing the antenna gain $G$-samples in the loss function with the PCE model, instead of utilizing the full-wave solver}. It is noted that this justification should be applicable to all other stochastic optimization algorithms that could be employed in lieu of the PSO.

\bibliographystyle{IEEEtran}
\bibliography{IEEEabrv,references.bib}

\end{document}